\documentclass[12]{article}

\renewcommand{\title}[1]{\begin{center}\bf\Large #1\end{center}}

\renewcommand{\author}[1]{\begin{center}\large #1\end{center}}

\usepackage{latexsym,amsfonts}                 
\newcommand{\rr}{\mathbb{R}}                   

\begin {document}                 

\title{Quantisation of Gauged $SL(2,\rr)$ WZNW Theories}

\author{
 George Jorjadze${}^a$ 
\footnote{Talk presented at the 35th International Symposium Ahrenshoop, 
Theory of Elementary Particles, August 26-31, 2002, 

\hspace{2mm}
Wernsdorf/Berlin, Germany.}
\footnote{email: \tt jorj@rmi.acnet.ge}
 and Gerhard Weigt${}^b$
\footnote{email: \tt weigt@ifh.de} \\
{\small${}^a$Razmadze Mathematical Institute,}\\
  {\small M.Aleksidze 1, 380093, Tbilisi, Georgia}\\
{\small${}^b$DESY Zeuthen, Platanenallee 6,}\\
{\small D-15738 Zeuthen, Germany}}

\begin{abstract}

\noindent
Canonical quantisation of the free-field zero modes $q$, $p$ on a
half-line $p>0$ provides for WZNW coset theories self-adjoint vertex operators
on account of hidden symmetries generated by an $S$-matrix.

\end{abstract}

\section{Introduction}

The cosets of the $SL(2,\rr)$ WZNW model, the Liouville theory as well
as the $SL(2,\rr)/\rr_+$ and $SL(2,\rr)/U(1)$ black hole models, are
classically completely described by gauge invariant Hamiltonian
reduction. This is reviewed in \cite{JW2}. A canonical quantisation
can be performed in the same way as it has been done for the Liouville
theory in \cite{OW, JW1}. Here one uses the general solution of the
coset as a canonical transformation between the non-linear coset
fields and free fields, replaces the Poisson brackets of the canonical
free fields by commutators, normal orders non-linear expressions in
the free fields, and avoids ensuing anomalies by deforming the
composite operators of the coset theories \cite {OW, JW1, FJW}.

Our aim is to construct vertex operators for the calculation of coset
correlation functions. However the coset dynamics restricts the
free-field zero modes $q$, $p$ to the half-line $p>0$ where the
coordinate operator $\hat q=i \hbar \partial_p$ is not self-adjoint.
We can expect  that the vertex operators which
contain $e^{\pm \gamma \hat q}$, nevertheless, act self-adjointly on
$L^2(\rr_+)$ due to  symmetries provided by $S$-matrix transformations of
the respective coset theory.

Quantum Liouville theory is the simplest example, and it proves to be
fundamental for an understanding of the quantisation of all the other
WZNW cosets.  It is therefore appropriate to restrict this short contribution
exclusively to the Liouville theory. Since its zero mode structure is
basically Liouville particle dynamics, we shall give first a full
description of the particle model. Guided by this structure we then 
describe the problem for the deformed field theoretical situation
\cite{JW3}, which refers to the Liouville oscillator vacuum only. It
might be worth mentioning here that the Liouville $S$-matrix 
 corresponds to a particular M\"obius transformation, which therefore leaves the
Liouville field invariant. The various cosets differ even in this
respect by technical peculiarities only.

\section{Liouville particle dynamics}

The Liouville vacuum configurations are described by
a homogeneous field $\partial_\sigma\varphi(\tau,\sigma)=0$ which becomes a 
time dependent coordinate $\varphi(\tau,\sigma)=Q(\tau)$. The 
Liouville equation
\begin{equation}\label{Liouville-equation}
\varphi_{\tau\tau}(\tau,\sigma) 
-\varphi_{\sigma\sigma}(\tau,\sigma)
+\frac{4m^2}{\gamma}\,\, e^{2\gamma\varphi(\tau,\sigma)}=0
\end{equation} 
reduces then to the mechanical model
\begin{equation}\label{m-Liouville}
\ddot{Q}(\tau) +\frac{4m^2}{\gamma}\, e^{2\gamma Q(\tau)}=0.
\end{equation}
Its Hamiltonian
\begin{equation}\label{m-Hamiltonian}
H=\frac{1}{4\pi}\left(P^2+4\omega^2\, e^{2\gamma Q}\right)
\end{equation}
is given by the canonically conjugate variables $Q$, $\,P=2\pi\dot{Q}$
and the parameter $\omega=2\pi m/\gamma$. 
The particle vertex function 
$V(\tau)=e^{-\gamma  Q(\tau)}$ satisfies the linear equation 
\begin{equation}\label{v-eq}
\ddot V(\tau) =\frac{\gamma^2}{\pi}\,H\, V(\tau),
\end{equation}
which can be integrated easily, and it provides 
the general solution of (\ref{m-Liouville})
\begin{equation}\label{g-solution}
e^{-\gamma Q(\tau)}=
e^{-\gamma \left(q+\frac{p\tau}{2\pi}\right)} +
\frac{\omega^2}{p^2}\,\,e^{\gamma \left(q+\frac{p\tau}{2\pi}\right)},
\end{equation}
with the (zero mode) integration constants $q$ and 
 \begin{equation}\label{p-H}
p=\sqrt {4\pi H}>0.
\end{equation}
The particle phase space coordinates $P,\,Q$ at $\tau=0$ and $p,\,q$ 
are related by a canonical and invertible map of the plane $(P,Q)$ onto the
half-plane $(p,q)$
\begin{eqnarray}\label{PQ-pq}
e^{-\gamma Q}=
e^{-\gamma q} +
\frac{\omega^2}{p^2}\,\,e^{\gamma q},~~~~~~~~~
P=-p\,\tanh \left(\gamma q+\log\frac{\omega}{p}\right).
\end{eqnarray}
Due to this half-plane situation and the non-linear character of this map 
a self-adjoint quantum realisation of the vertex (\ref{g-solution})
becomes non-trivial.

The mechanical Liouville model is asymptotically a 
free theory and (\ref{g-solution}) gives 
\begin{equation}\label{in-varibales}
\lim_{\tau \rightarrow -\infty}\,
\left[Q(\tau) - q-\frac{p\tau}{2\pi}\right]=0,
~~~~~~~~~~\lim_{\tau \rightarrow -\infty}\,\left[ P(\tau)-p\right] =0.
\end{equation}
So $p$, $\,q$ can be interpreted as the $in$-variables of 
Liouville particle dynamics, and the half-plane condition $p>0$ is
indeed consistent with the positive $in$-momentum.

Similarly $out$-variables can be defined by (\ref{g-solution}) as 
$\tau \rightarrow +\infty$. The $in$- and $out$-variables are related 
by the transformation
\begin{equation}\label{out-varibales}
P_{out}=-p, ~~~~~~Q_{out}=-q+\frac{2}{\gamma}\log\,\frac{p}{\omega}\,,
\end{equation}
which combines the reflection of $p,\,q$ with a simple canonical map.
It is an important observation that the general solution
(\ref{g-solution}) is invariant under the transformation
(\ref{out-varibales}). Quantum mechanically this symmetry will be
generated by the $S$-matrix of the particle theory (\ref{s_k}). In
Liouville field theory $in-$ and $out-$fields \cite{J} are related by
the special M\"obius transformation $A\rightarrow -(m^2\,A)^{-1},~~
\bar A\rightarrow -(m^2\,\bar A)^{-1}$, which keeps invariant the
general solution
\begin{equation}\label{Liouville-solution}
\varphi(\tau,\sigma) =\log \frac{A\,'(\tau +\sigma)\bar A\,'(\tau -\sigma)}
{[1+m^2 A(\tau +\sigma)\bar A(\tau -\sigma)]^2}\,.
\end{equation}

\section{Vertex operator}

To obtain the $\hat p$, $\hat q$ operator structure of a vertex operator,
we quantise first the mechanical Liouville model in the
$Q$-representation.  It will be convenient to use the notation
\begin{equation}\label{alpha}
x=\gamma Q-\log\frac{\alpha}{m}~~~~\mbox{and}~~~~ 
\alpha =\frac{\hbar\gamma^2}{4\pi}\,,
\end{equation}
in which the operator for the Hamiltonian (3) becomes
\begin{equation}\label{H,H_0}
\hat H=\hbar\alpha\left(-\partial^2_x+e^{2x}\right).
\end{equation}
The eigenstates of this Hamiltonian are solutions of the
Schr\"{o}dinger equation
\begin{equation}\label{H-Psi}
-\Psi_k\,''(x)+ e^{2x}\,\Psi_k(x)=k^2\,\Psi_k(x).
\end{equation}
They are given by Kelvin functions $K_{ik}\left(e^{x}\right)$ which
are real, and $K_{-ik}(e^x)=K_{ik}(e^x)$ \cite{DJ}.  
(To simplify formulas we use here $k$ instead of $p=\hbar\gamma k$.)
The spectrum $E=\hbar\alpha k^2$ is non-degenerate and 
it does not contain the point $k=0$.
We take therefore the eigenstates
$\Psi_{k}(x)$ with $k>0$ only, which reflects the half-plane situation
quantum mechanically.

Choosing the normalisation 
\begin{equation}\label{d_k}
\Psi_k(x)=d_k \,K_{ik}\left(e^{x}\right),~~~~~\mbox{with}~~~~
d_k=\sqrt{\frac{2}{\pi}}\,\,\frac{2^{ik}}{\Gamma(-ik)}\,,
\end{equation}
the eigenstates will have the asymptotic behaviour
\begin{equation}\label{Psi-}
\Psi_k(x)\rightarrow \frac{e^{ikx}}{\sqrt{2\pi}}+
S_k\,\frac{e^{-ikx}}{\sqrt{2\pi}}\,,~~~~\mbox{as}~~~x\rightarrow -\infty,
\end{equation}
with the reflection amplitude \cite{BCGT} 
\begin{equation}\label{s_k}
S_k=2^{2ik}\,\frac{\Gamma(ik)}{\Gamma(-ik)}\,.
\end{equation}

We are now able to calculate matrix elements of an arbitrary vertex
operator $e^{2b\,\gamma\hat Q(\tau)}$ between the eigenstates of (\ref{H,H_0}) 
(Please, note that we also
use the notation $V$ for $V_{-1}$$\,$!) 
\begin{equation}\label{V(k,k)}
V_{2b}(k,k\,';\tau)=\langle\,\Psi_k |e^{2b\,\gamma\hat Q(\tau)}
|\,\Psi_{k\,'}\,\rangle.
\end{equation}
The result
\begin{equation}\label{e^Q(tau)1}
V_{2b}(k,k\,';\tau)=\left(\frac{\alpha}{m}\right)^{2b}\,d_k^*\,d_{k\,'}
\,\,e^{i\alpha(k^2-k\,'^2)\tau}\,\,F_{2b}(k,k\,'),
\end{equation}
can be expressed by the integral
\begin{equation}\label{I(k,k)}
F_{2b}(k,k\,')=
\int_{-\infty}^{\infty}dx\, K_{ik}(e^x)\,e^{{2b} x}\,K_{ik\,'}(e^x).
\end{equation}
But this integral is well defined for $b >0$ only, 
and  it diverges  for  $b \leq 0$. 
Using the notation
\begin{equation}\label{kappa}
\kappa= \frac{k+k\,'}{2}\,\,,~~~~~~~\rho=\frac{k-k\,'}{2}\,\,,
\end{equation}
for $b>0$ the integration yields \cite{BCGT}
\begin{equation}\label{V(k,k)1}
V_{2b}=\left(\frac{\alpha}{m}\right)^{2b}\,\,e^{4i\alpha\rho\kappa\tau}\,4^{b-i\rho}\,
\frac{\Gamma\left(b+i\kappa\right)
\Gamma\left(b-i\kappa\right)}
{\Gamma\left(i\rho+i\kappa\right)
\Gamma\left(i\rho-i\kappa\right)}\,\frac
{\Gamma\left(b+i\rho\right)
\Gamma\left(b-i\rho\right)}{4\pi\,\Gamma(2b)}\,.
\end{equation}
In order to get the vertex function for negative $b$ one needs a
smooth continuation of (\ref{V(k,k)1}) as a generalised function.
Since $\kappa>|\rho|$ ambiguities can arise at $b=-n$,
$n=0,1,2,...\,$. Near these points if $b=-n+\epsilon$ (\ref{V(k,k)1})
behaves like
\begin{equation}\label{delta+}
\frac{1}{\pi}\,\frac{\epsilon}{(k-k\,')^2+\epsilon^2}.
\end{equation}
One can show that this generalised function has for holomorphic test
functions the following smooth continuation from positive to negative
values of $\epsilon$ \cite {JW3}
\begin{equation}\label{delta-}
\frac{1}{\pi}\,\frac{\epsilon}{(k-k\,')^2+\epsilon^2}+
\delta(k-k\,'+ i\epsilon)+\delta(k-k\,'- i\epsilon),
\end{equation}
where the $\delta$-function with complex arguments 
is defined in the standard manner 
\begin{equation}\label{delta}
\int_0^{+\infty} dk\,'\,\,\delta(k-k\,'\pm i\epsilon)
\,\psi (k\,')=\psi (k \pm i\epsilon).
\end{equation}
The continuation of (\ref{V(k,k)1}) to negative values of $b$ so
creates a pair of $\delta$-functions with complex arguments each time
$b$ passes a negative integer value, and for $b=-|b|$ results  
\begin{eqnarray}\label{V(k,k)2}
V_{-2|b|}=V_{-2|b|}^{+}\,(\kappa,\rho;\tau)
+~~~~~~~~~~~~~~~~~~~~~~~~~~~~~~~~~\nonumber\\
\left(\frac{m}{\alpha}\right)^{2|b|}\,\sum_{l=0}^{[|b|]}\, C_{2|b|}^l 
e^{-4\alpha(|b|-l)\kappa\tau}\,\,
\frac{\Gamma(-|b|+i\kappa)\,\Gamma(-|b|-i\kappa)}
{4^l\,\,\Gamma(-|b|+l+i\kappa)\,\Gamma(-|b|+l-i\kappa)}
\,\delta[2\rho-2i(|b|-l)] \nonumber\\
+\, C_{2|b|}^l\,e^{4\alpha(|b|-l)\kappa\tau}\,\,
\frac{\Gamma(-|b|+i\kappa)\,\Gamma(-|b|-i\kappa)}
{4^{2|b|-l}\,\,\Gamma(|b|-l+i\kappa)\,\Gamma(|b|-l-i\kappa)}
\,\delta[2\rho+2i(|b|-l)].
\end{eqnarray}
Here $V_{-2|b|}^{+}\,(\kappa,\rho;\tau)$ is defined by the r.h.s of
(\ref{V(k,k)1}) where $b$ is replaced by $-|b|$, $ [|b|]$ is
the integer part of $|b|$, and
\begin{equation}\label{C_b}
C_{2|b|}^l=\prod_{j=0}^{l-1}\frac{2|b|-j}{j+1}.
\end{equation}
In particular, the function $V_{-2|b|}^{+}$ vanishes at half-integer
$|b|$ due to the pole of $\Gamma(-2|b|)$, and it creates at integer
$|b|$ the terms proportional to $\delta(\rho)$, so that for $2|b|=n$
one has
\begin{equation}\label{V(k,k)3}
V_{-n}(\kappa,\rho;\tau)=\left(\frac{m}{\alpha}\right)^{n}\sum_{l=0}^nC_n^l\,
e^{-2(n-2l)\kappa\tau}\,\,\prod_{j=0}^{l-1} \frac{1}{4\kappa^2+(n-2j)^2}
\,\,\delta[2\rho-i(n-2l)],
\end{equation}
where $C_n^l$ are now binomial coefficients. 

Our aim is to get the vertex operator in terms of the zero mode operators
$\hat p$, $\hat q$. So we have to establish a connection between the
$Q$-representation and the standard $p$-representation on the half-plane. 
Wave functions of the
$p$-representation are $\psi(p)\in L^2(\rr_+)$ and the operators
$e^{\pm\gamma\hat q}$ can be defined as
\begin{equation}\label{e^}
e^{\pm\gamma\hat q}=e^{\pm(i\hbar\gamma\partial_p\,+\lambda)},
\end{equation}
with some constant $\lambda$. Due to (\ref{p-H}) the eigenstates of
(\ref{H,H_0}) $|\,\Psi_k\,\rangle$ can also be identified with the
eigenstates of the momentum operator $\hat p\,
|\,\Psi_k\,\rangle=\hbar\gamma \,k|\,\Psi_k\,\rangle$.  This
identification connects the wave functions $\psi(p)$ with wave
functions of the $Q$-representation $\Psi(x)$ by
\begin{equation}\label{p-Q}
\psi(p) =\int_{-\infty}^{+\infty} dx\,\, 
\Psi_k^*(x)\,\Psi(x),~~~~~\mbox{for}~~~~~
 p=\hbar\gamma \,k.
\end{equation}
Since the eigenstates $|\Psi_k\rangle$ are complete
this transformation is unitary.

Let us now consider as the simplest non-trivial example the
vertex operator $\hat V(\tau)$.
Its matrix elements  correspond to the case $n=1$ of (\ref{V(k,k)3})
\begin{equation}\label{V(k,k,tau)}
V(k,k\,';\tau)=\frac{m}{\alpha}\,\left(
e^{-\alpha(k+k\,')\tau}\,\,\delta(k-k\,'-i)+
\frac{e^{\alpha(k+k\,')\tau}}{4\,k\,k\,'}\,\,\delta(k-k\,'+i),
\right). 
\end{equation}
From this expression one can easily read off the zero mode operator structure
\begin{equation}\label{e^Q=e^q}
\hat V(\tau)=
e^{-\frac{\gamma\hat p}{4\pi}\tau}\,\, 
e^{-\gamma \hat q}\,\,e^{-\frac{\gamma\hat p}{4\pi}\tau}+
\omega^2\,\frac{e^{\frac{\gamma\hat p}{4\pi}\tau}}{\hat p}\,\,
e^{\gamma \hat q}\,\,\frac{e^{\frac{\gamma\hat p}{4\pi}\tau}}{\hat p}.
\end{equation}
This fixes the parameter $\lambda$ of (\ref{e^}) as
$\,\lambda=\log(\frac{\alpha}{m})$.  One can show that the vertex
operator $\hat V_{-n}$ is the $n$-th power of (\ref{e^Q=e^q}), but for
positive or non-integer negative $2b$ $\hat V_{2b}$ is given by an
infinite series of $q$-exponentials, much as in quantum Liouville
field theory \cite{OW, JW3}.

It remains to consider the self-adjoint action of the vertex operator
(\ref{e^Q=e^q}) on $L^2(\rr_+)$. Since asymptotically for
$\tau\rightarrow\pm\infty$ only one term survives one is tempted to
demand hermiticity for each term of (\ref{e^Q=e^q}) separately.
But a proof of hermiticity for $e^{\gamma \hat q}$ would require very
special boundary conditions on holomorphic
functions $\psi(p)$ at $\mbox{Re}\,\, p=0$, a mathematically by itself
interesting but still unsolved problem \cite{JW3}.

However, the vertex operator (\ref{e^Q=e^q}) can be shown to become
self-adjoint as a whole \footnote{We thank an anonymous referee of
  \cite{JW3} for a simple realisation of this idea. 
In fact this idea 
was intensively discussed in \cite{T}, but the results were inconclusive.}
on account of its symmetry under
transformations given by the $S$-matrix
\begin{equation}\label{S}
\hat S = \hat{\cal{P}}S(p).
\end{equation}
Here $S(p)$ is the multiplicative reflection amplitude (\ref{s_k})
with $p=\hbar\gamma k$ and
$\hat{\cal{P}}$ the parity operator $\hat{\cal{P}}\psi(p)=\psi(-p)$.
It is easy to see that (\ref{S}) replaces the $in$-coming first term of 
(\ref{e^Q=e^q}) by the $out$-going second one, and vice versa,
so that the particle vertex operator remains invariant.
That means $\hat V_{out} (\tau)=\hat S \hat V (\tau)\hat S^{-1}$ is
identical to (\ref{e^Q=e^q}). This also holds in general as one can see 
from (\ref{V(k,k)1}) and (\ref{V(k,k)2}). 
 The $S$-matrix
is just the quantum version of the symmetry transformation
(\ref{out-varibales}).
$\hat S$ also maps the Hilbert space of the $in$-fields
$L^2(\rr_+)$ onto $L^2(\rr_-)$ for the $out$-fields, which for the wave 
functions $\psi(p)\in L^2(\rr_+)$, $\tilde\psi(p)\in L^2(\rr_-)$
is given by $\tilde\psi(-p)=S(p)\psi(p)$. The last relation is defined by 
(\ref{p-Q}) using (\ref{d_k}) and $\Psi_{-k}^*(x)=
d_{-k}^*\,K_{-ik}=S_{k}\,\Psi_{k}^*(x)$.

 Due to these properties one can extend the definition
of the matrix element of the vertex operator from $L^2(\rr_+)$ to $L^2(\rr)$
\begin{equation}\label{R+R}
\int_0^\infty dp\,\,\psi_+^*(p)\hat V(\tau)\psi_+(p)=\frac{1}{2}
\int_{-\infty}^\infty dp\,\,\Psi^*(p)\hat V(\tau)\Psi(p).
\end{equation}
Here $\Psi(p)=\psi(p)$ for $p>0$,
$\Psi(p)=\tilde\psi(p)=S(-P)\psi(-p)$ for $p<0$, and $\Psi(p)\in
L^2(R)$ satisfies $\hat S\Psi(p)=\Psi(p)$.  Self-adjointness of the
operator (\ref{e^Q=e^q}) obviously holds on $L^2(\rr)$ for holomorphic
wave functions $\Psi(p)$. As a consequence, self-adjointness of $\hat
V(\tau)$ on $L^2(\rr_+)$ requires for $\psi(p)$ a holomorphic
extension to the negative half-line so that
$\psi(-p)=\tilde\psi(-p)=S(p)\psi(p).$ Such functions are given by
$\psi(p)=d\,^*(p) f(p)$, where $d(p)$ is defined by (\ref{d_k}) with
$p=\hbar\gamma k$, $f(p)$ is an even holomorphic function
$f(-p)=f(p)$, and (\ref{R+R}) is well defined at $p=0$ since $\psi(0)=0$
on account of $d(0)=0$.

\section{Reduced Liouville field theory}

The Liouville vertex operator which corresponds to (\ref{e^Q=e^q}) results as
an oscillator vacuum matrix element \cite{OW, JW3}
\begin{equation}\label{V(bb)}
{\bf \hat V}(\tau)
=e^{-\frac{\gamma\hat p}{4\pi}\tau}\,\,
e^{-\gamma \hat q}\,\,e^{-\frac{\gamma\hat p}{4\pi}\tau}+
\omega_\alpha^2\,\,\frac{e^{\frac{\gamma\hat p}{4\pi}\tau}}{\hat p}\,\,   
\frac{\Gamma\left(-i\frac{\gamma p}{2\pi}\right)}
{\Gamma\left(i\frac{\gamma p}{2\pi}\right)}\,\,
e^{\gamma \hat q}\,\,
\frac{\Gamma\left(i\frac{\gamma p}{2\pi}\right)}
{\Gamma\left(-i\frac{\gamma p}{2\pi}\right)}
\,\,\frac{e^{\frac{\gamma\hat p}{4\pi}\tau}}{\hat p}.
\end{equation}
where 
\begin{equation}\label{omega_alpha}
\omega_\alpha=\frac{2\pi m_\alpha\Gamma(1+2\alpha)}{\gamma}~~~~
\mbox{and}~~~~~m_\alpha^2 =\frac{\sin 2\pi\alpha}{2\pi\alpha}\,\,m^2\, . 
\end{equation}
Both, the parameter $\omega_\alpha$ and the $\Gamma$-functions
describe classical-valued quantum deformations which are due to
oscillator as well as zero mode contributions.

Asking the question whether the operator (\ref{V(bb)}) 
is symmetric under  a $S$-matrix transformation $\hat S_L=\hat{\cal P}S_L(p)$
which exchanges the $in$-coming with the deformed $out$-going zero mode part,
we found as an answer the deformed reflection amplitude
\begin{equation}\label{S1}
S_L(p) =-\left(\frac{4\pi m\alpha^2}{\Gamma(1+2\alpha)\,\sin 2\pi\alpha }
\right)^{2ik}
\,\frac{\Gamma\left(i\frac{\gamma p}{4\pi\alpha}\right)}
{\Gamma\left(-i\frac{\gamma p}{4\pi\alpha}\right)}\,\
\frac{\Gamma\left(i\frac{\gamma p}{2\pi}\right)}
{\Gamma\left(-i\frac{\gamma p}{2\pi}\right)}\,.
\end{equation}
This reflection amplitude is deformed in comparison with the particle
 case (\ref{s_k}), and it is surprisingly identical to that derived in
 \cite{ZZ} by a symmetry of a 3-point correlation function suggested
 in \cite{DO} on account of path-integral considerations.  This result
 motivates one to look for a deeper understanding of the path-integral
 conjectures by means of the operator approach, as discussed in
 \cite{T}.

For a proof of self-adjointness knowledge of the full $S$-matrix
for Liouville field theory is required.

\vspace{8mm}
\noindent
{\bf Acknowledgement}\\ G. Jorjadze thanks the organisers of the 
Ahrenshoop Symposium for financial support. His
research was supported by grants from the DFG, INTAS, RFBR and GAS.

\end{document}